\documentstyle[12pt,prl,aps,multicol,epsfig]{revtex}
\begin{document}
\author{K.L. Sebastian}
\address{Department of Inorganic and Physical Chemistry\\
Indian Instittue of Science\\
Bangalore 560012\\
India}
\title{Pulling a polymer out of a potential well and the mechanical unzipping of DNA%
\thanks{%
Dedicated to the memory of Professor A.K. Chandra, teacher and friend}}
\maketitle

\begin{abstract}
Motivated by the experiments on DNA under torsion, we consider the problem
of pulling a polymer out of a potential well by a force applied to one of
its ends. If the force is less than a critical value, then the process is
activated and has an activation energy proportinal to the length of the
chain. Above this critical value, the process is barrierless and will occur
spontaneously. We use the Rouse model for the description of the dynamics of
the peeling out and study the average behaviour of the chain, by replacing
the random noise by its mean. The resultant mean-field equation is a
nonlinear diffusion equation and hence rather difficult to analyze. We use
physical arguments to convert this in to a moving boundary value problem,
which can then be solved exactly. The result is that the time $t_{po}$
required to pull out a polymer of $N$ segments scales like $N^2$. For models
other than the Rouse, we argue that $t_{po}\sim N^{1+\nu }$
\end{abstract}


\section{Introduction}

\begin{figure}
 \epsfbox[200 100 700 700]{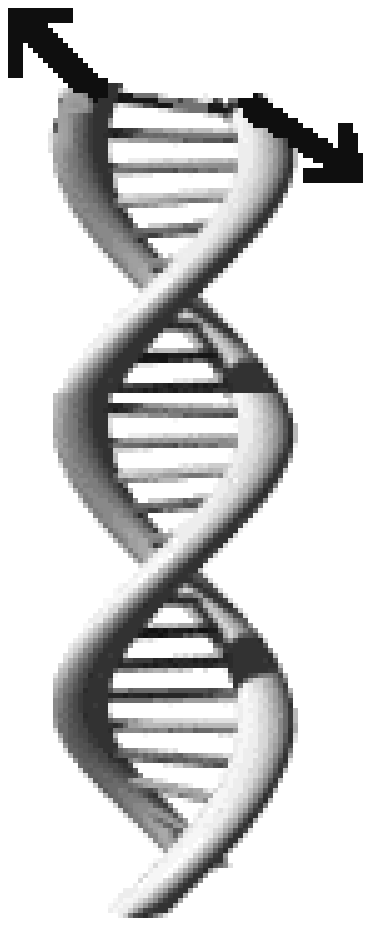}
\caption{The forced unzipping of DNA.}  \label{fig1}
\end{figure}

Considerable attention has been paid to the replication of the DNA molecule 
\cite{Kornberg}. The first step in the replication is the unzipping of the
two strands. Usually, this is caused by enzymes and mechanical force at the
molecular level is involved in the action of these enzymes. Recently,
micromanipulation techniques have been developed to study single molecules
of DNA under stress\cite{Austin}. Most such single molecule experiments have
been on the stretching properties of DNA, but very recently, the response of
the DNA to external torques has also been studied \cite{Strick,Leger}.
Mechanical unzipping of the two strands of the DNA, in the absence of
enzymes has been carried out \cite{Roulet}. There have been interesting
investigations in to the theory of stretching the DNA \cite{Marko} and its
denaturation under a torque (see Fig. \ref{fig1}) \cite{Bhattacharjee,Cocco}%
. In an interesting recent e-print, Bhattacharjee \cite{Bhattacharjee} has
suggested a minimal model to study the unzipping. He treats the DNA as
consisting of two flexible interacting elastic strings which are bound
together by an attractive interaction. The two are tied together at one end
and at the other end, there are forces acting on the two strands, trying to
separate them. Winding is ignored in his analysis. He assumes that in
relative coordinates, the Hamiltonian for the unzipping problem is
equivalent to that of for a single chain, subject to an attractive short
range potential well with a force acting at one of its ends. It is shown
that the polymer can be pulled out only if the force exceeds a critical
value. The model is of interest in other contexts too. For example, one can
imagine constructing micro-fabricated channels having different depths in
different regions to trap the polymer by entropic barriers (see the recent
experiments by Han et. al. \cite{Han}) and then pulling it out from the
traps by the application of a force at one end. Another possible application
is the desorption of a long chain molecule adsorbed on a surface by pulling
at one end. Our aim in this paper is to investigate the dynamics of the
escape of a chain from a potential well as a result of applying a force at
one of its ends. We consider only the one dimensional problem and our
approach is much less sophisticated than that of \cite{Bhattacharjee}, as we
do not take in to account the excluded volume interaction. We bring out the
physical reason for the existence of the critical force. For describing the
dynamics, we use the simple Rouse model. We find that $t_{po}$, the time
required to pull out a chain molecule of $N$ segments scales as $N^2$.

In the section ~\ref{statmech} we give an outline of the statistical
mechanics of a polymer, trapped by an attractive potential, with a force
acting at one end. Section \ref{Rouse} outlines the Rouse model that is
applicable for a polymer subject to a force, and analyze why the force has
to exceed a critical force for pulling it out. In section \ref{dynamics}, we
argue that the problem can be approximated by a moving boundary value
problem and then solve the problem. Finally, section \ref{conclusions}
summarizes our conclusions.

\section{Statistical Mechanics of a trapped polymer with a force at one of
its ends}

\label{statmech}

\begin{figure}
\epsfbox[200 200 500 500]{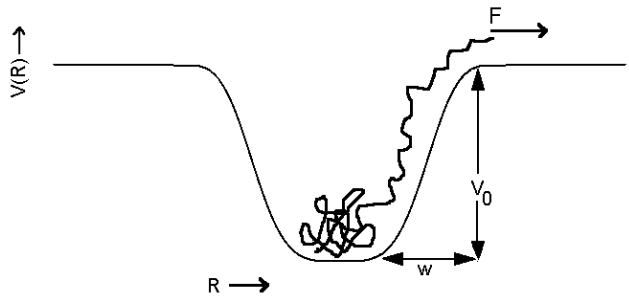}
\caption{The polymer trapped in a potential well, subject to a force $F$.}
 \label{fig2}
\end{figure}

In this section we consider the statistical mechanics of a polymer in one
dimension, trapped in an attractive potential $V(R)$, that is located near $%
R=0$ (see \ref{fig2} for the nature of the potential) and subject to a force 
$F$ at one of its ends. We take the number of segments in the polymer to be $%
N$. The partition function for a free polymer, that is not subject to any
force, but having one end at $R=0$ may be calculated as $Z=\int dRG(R,0,N)$
where $G(R,0,N)$ is the propagator for the unforced molecule, which obeys
the differential equation $(\frac \partial {\partial N}-\frac{l^2}6\frac{%
\partial ^2}{\partial N^2}+\frac 1{k_BT}V(R))G(R,R_0,N)=\delta (N)\delta
(R-R_0)$ (see \cite{DE}). $l$ is the Kuhn length for the chain and $R$ and $%
R_0$ denote the positions of the two ends of the chain.

In the presence of a force $F$ acting at the end at $R$, the partition
function becomes

\begin{equation}
\label{sec1one}Z=\int_{-\infty }^\infty dRG(R,0,N)e^{\frac{RF}{k_BT}} 
\end{equation}
Using the ground state dominance approximation for the propagator, valid for
long polymers \cite{DE}, we can write the above as:

\begin{equation}
\label{sec1two}Z=\int_{-\infty }^\infty dR\psi _0^{*}(R)\psi _0(0)e^{\frac{%
RF-E_0N}{k_BT}}
\end{equation}
where $\psi _0(R)$ is the lowest eigen-function of the operator $H=-\frac{l^2%
}6\frac{\partial ^2}{\partial R^2}+\frac 1{k_BT}V(R)$, having the eigenvalue 
$E_0/(k_BT)$. Now, for large values of $R$, the eigen-function $\psi
_0(R)\sim $ $\exp (-\sqrt{\frac{-6E_0}{l^2k_BT}}R)$ $\sim \exp (-\sqrt{\frac{%
6V_0}{l^2k_BT}}R)$, where in the last step we have neglected the ''zero
point energy'' contribution to the lowest eigenvalue and approximated it by $%
-V_0$, where $V_0$ is the depth of the potential well.  On using this in the
equation (\ref{sec1two}), we realize that for $F>F_c=\sqrt{\frac{6k_BTV_0}{%
l^2}}$, the contribution from large values of $R$ make the integral diverge,
indicating that for $F>F_c$, the polymer gets pulled out of the hole. For $%
F<F_c$, the partition function is finite, indicating that the molecule
remains trapped. In this analysis, we have neglected the zero-point energy,
which in the polymer problem, is equivalent to neglecting the effect of
thermal fluctuations. Thermal fluctuations make $F_c$ slightly lower \cite
{Bhattacharjee}.

\section{The Rouse Model and the free energy profile for the pulled polymer}

\label{Rouse}

\begin{figure}
  \epsfbox[200 200 540 500]{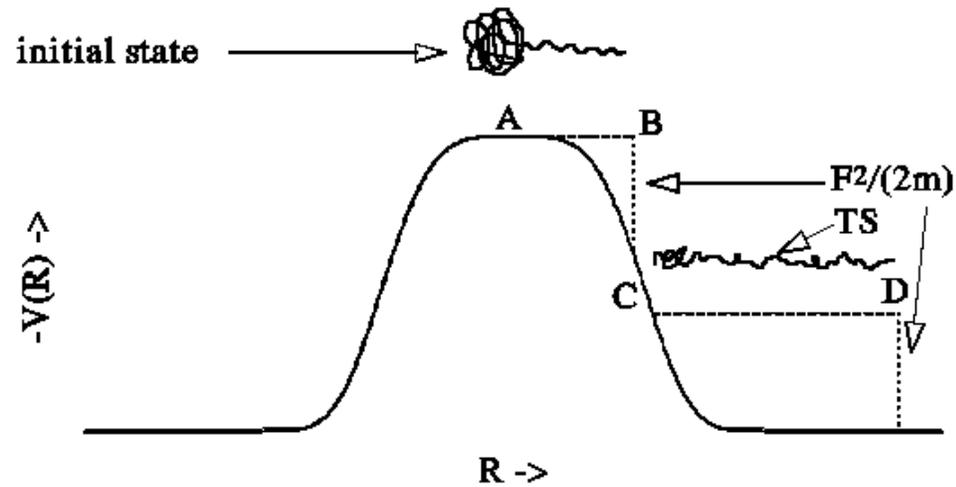}
\caption{The upside down potential $-V(R)$. The lines AB and CD represent the solutions
 of the Eq. \ref{one5} corresponding to the initial and the transition states respectively.  
The corresponding configurations of the polymer are represented by the coiled curves above them.  TS stand for transition state. The kinetic energy at the $n=0$ end is $\frac{F^{2}}{2 m}$ and this must be less than the height of the potential $V_{0}$, if the extrema are to exist.}
 \label{fig3}
\end{figure}

\begin{figure}
\vspace{6in}
  \epsfbox[200 200 500 500]{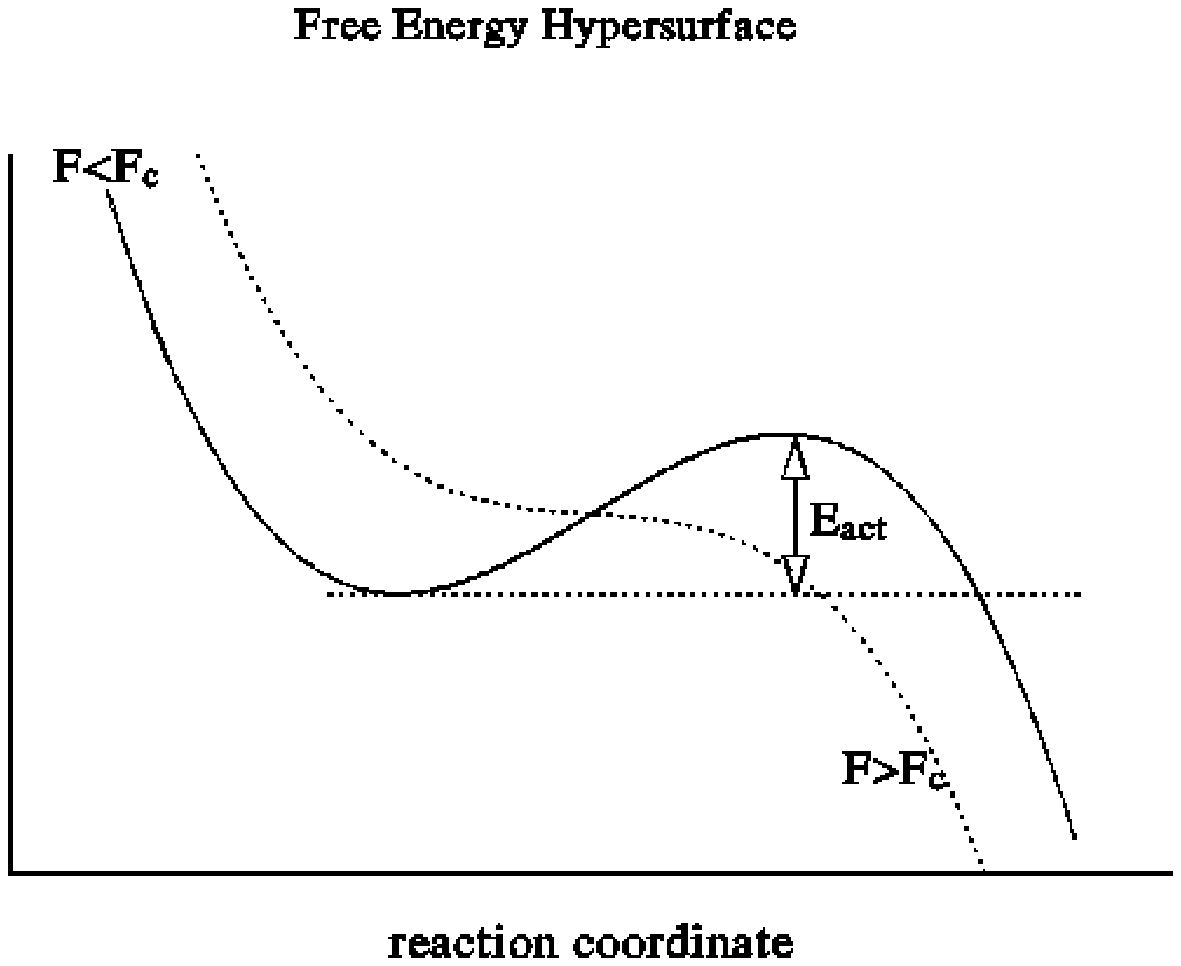}
\caption{The free energy hypersurface.  For $F<F_{c}$ it has both a local maximum and
 a minimum, while for $F>F_{c}$ no local extrema exist.  $E_{act}$ is the activation energy and is equal to $\frac{N\left( F_c^2-F^2\right) }{2m}$. } \label{fig4}
\end{figure}

The process that we study is shown in the Fig. \ref{fig2}. For its
description,we use the continuum limit of the Rouse model, discussed in
detail by Doi and Edwards \cite{DE}, which has recently been used to study
the related problem of escape of a polymer over a barrier \cite{Sebastian}.
The chain is approximated as a string, with segments (beads) labelled by
their position $n$ along the chain. $n$ is taken to be a continuous
variable, having values ranging from $-N$ to $0$ (this is convenient for the
present problem). The position of the $n^{th}$ segment in space (one
dimension) is denoted by $R(n,t)$, where $t$ is time. In the Rouse model,
this position undergoes overdamped Brownian motion and its time development
is described by the equation

\begin{equation}
\label{one}\zeta \frac{\partial R(n,t)}{\partial t}=m\frac{\partial ^2R(n,t)%
}{\partial n^2}-V^{\prime }(R(n,t))+f(n,t). 
\end{equation}
In the above, $\zeta $ is a friction coefficient for the $n^{th}$ segment.
The term $m\frac{\partial ^2R(n,t)}{\partial n^2}$ comes from the fact that
stretching the chain can lower its entropy and hence increase its free
energy. Consequently, the parameter $m=3k_BT/l^2$ (see Doi and Edwards \cite
{DE}, Eq. (4.5). They use the symbol $k$ for the quantity that we call $m$)
. $V(R)$ is the free energy of a segment of chain, located at the position $%
R $ and $V^{\prime }(R)=\partial V(R)/\partial R$. $f(n,t)$ are random
forces acting on the $n^{th}$ segment and have the correlation function $%
\left\langle f(n,t)f(n_1,t_1)\right\rangle =2\zeta k_BT\delta (n-n_1)\delta
(t-t_1)$ (see \cite{DE}, Eq. (4.12)). The deterministic part of the Eq. (\ref
{one}), which will play a key role in our analysis, is obtained by replacing
the random noise term in (\ref{one}), by its mean. As we are applying a
force at one end of the polymer (at $n=0$) and as the other end is free, the
boundary conditions to be satisfied are 
\begin{equation}
\label{one1}\left\{ \frac{\partial R(n,t)}{\partial n}\right\} _{n=0}=\frac F%
m 
\end{equation}

\begin{equation}
\label{one2}\left\{ \frac{\partial R(n,t)}{\partial n}\right\} _{n=-N}=0. 
\end{equation}
The above equations may equivalently be written as

\begin{equation}
\label{one3}\zeta \frac{\partial R}{\partial t}=-\frac \delta {\delta R(n)}%
E[R)]+f(n,t). 
\end{equation}
where we have defined the free energy functional

\begin{equation}
\label{one4}E[R]=\int_{-N}^0dn\left\{ \frac m2\left( \frac{\partial R}{%
\partial n}\right) ^2+V(R)-FR(n)\delta (n+\varepsilon )\right\} 
\end{equation}
$\varepsilon $ ($\rightarrow 0$), is a small positive number. The process of
pulling out may be made physically clearer by analyzing this free energy
functional. For this, we consider the free energy hypersurface that results
from the Eqs. (\ref{one1}) (\ref{one2}) and (\ref{one4}). We find the
extrema on this surface by putting $\frac \delta {\delta R}E[R]=0$, which
leads to

\begin{equation}
\label{one5}m\frac{\partial ^2R}{\partial n^2}=V^{\prime }(R)\text{ for }n<0 
\end{equation}
This has to be solved, subject to the conditions of the Eqs. (\ref{one1})
and (\ref{one2}). Imagining and speaking of $n$ as time for the rest of this
section, we realize that this is just Newton's equation for a particle of
mass $m$, moving in the upside down potential $-V(R)$. In the Fig. \ref{fig3}%
, we show this potential. The boundary condition of Eq. (\ref{one2}) means
that the particle has to start at the time $n=-N$, with zero velocity and
end at the time $n=0$, with the velocity $F/m$. This means that at the end
of its path, its kinetic energy is $\frac{F^2}{2m}$. At $n=-N$, the particle
can at the most start at the top of the potential, where its total energy is 
$V_0$. If it obeyed the equation (\ref{one5}), then at $n=0$, its kinetic
energy has to be less than this amount. Thus we get the condition that $V_0>%
\frac{F^2}{2m}$ (or equivalently, $F_c>F$) for the free energy surface to
have a local minimum). If this condition is satisfied, then for any finite
value of $N$, we can find two solutions to the equation (\ref{one5})
satisfying the boundary conditions (this is true for potentials of the shape
shown in the figure). These two solutions are shown in the Fig. \ref{fig3}.
In the first solution (shown as line AB in the figure), the particle starts
near the maximum of $-V(R)$, at the time $n=-N$ and takes a very long time
to move away from the maximum, and eventually arrives at the point B at the
final time $n=0$ with exactly the right kinetic energy of $\frac{F^2}{2m}$.
As this corresponds to a configuration for the polymer where almost all its $%
N$ (assumed to be large) segments are near the minimum of the potential well 
$V(R)$, we can estimate the free energy for this configuration to be $\sim
-NV_0$. The second solution, also shown in the Fig. \ref{fig3}, CD,
corresponds to the particle starting from rest from an appropriate value of $%
R$ (point C in the Fig. \ref{fig3}) at the time $n=-N$ and arriving at $n=0$
with the kinetic energy $\frac{F^2}{2m}$. The particle spends only
relatively small amount of time in the region where the potential energy $%
-V(R)$ is non-zero. One easily estimates the free energy of the polymer to
be $\sim -N\frac{F^2}{2m}$. Thus, for a polymer of $N$ segments, there is a
barrier of height $N(V_0-\frac{F^2}{2m})=N(F_c^2-F^2)/2m$, to be overcome
for the pulling the polymer out. As $F\rightarrow F_c$ this barrier is
reduced and if $F>F_c$ then the free energy surface has no local minimum -
its minimum is at infinity - this means that the pulling out is barrierless
and would happen spontaneously. A pictorial representation of this scenario
is given in the Fig. \ref{fig4}.

\section{The Dynamics}

\label{dynamics}

We now consider the case with $F>F_c$ so that the pulling out of the chain
has no activation barrier. We are interested in only the average behavior of
the segments, in the process of pulling out. This may be obtained by
analyzing the deterministic equation

\begin{equation}
\label{two}\zeta \frac{\partial R(n,t)}{\partial t}=m\frac{\partial ^2R(n,t)%
}{\partial n^2}-V^{\prime }(R(n,t)) 
\end{equation}
where we have replaced the random term $f(n,t)$ by its average value, viz., $%
0$. We now wish to analyze this equation subject to the two boundary
conditions of the Eqs. (\ref{one1}) and (\ref{one2}). This equation, is just
a nonlinear diffusion equation.

\begin{figure}
\vspace{6in}
  \epsfbox[200 200 590 500]{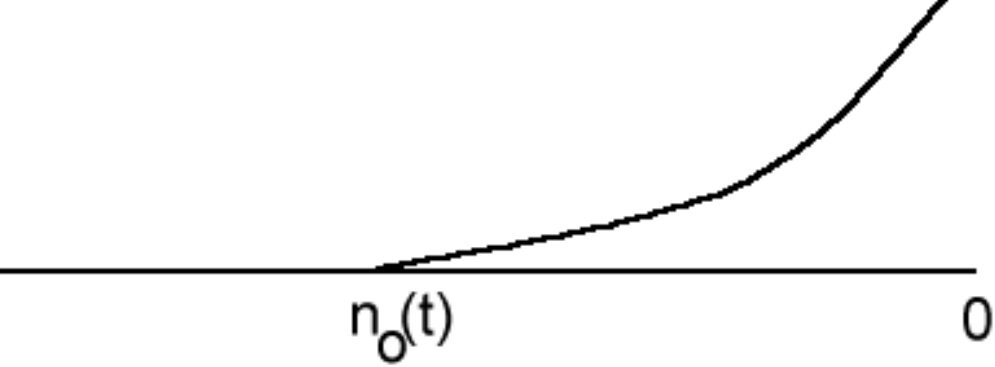}
\caption{The moving boundary value problem.  The force $F$ is applied at $n=0$.  
$n$ varies from $-N$ to $0$.  At a given time $t$, the portion with $0 > n > n_{o}(t)$ 
has been pulled out from the well.}  \label{fig5}
\end{figure}

The equation (\ref{two}) is very difficult to solve for any realistic $V(R)$%
. However, we can use the following physical picture. We take $V(R)$ to be a
short ranged well having a depth of $V_0$, and to have a rather sharp
boundary like the one in the figure \ref{fig2}. The change over from $-V_0$
to the flat region where $V(R)=0$ occurs over a region of width $w$, which
is assumed to be small. This means that $V^{\prime }(R(n,t))$ resembles a
Dirac delta function, which implies that $\frac{\partial R(n,t)}{\partial n}$
has to change very rapidly, across this region, and would resemble a step
function. If we consider the process of escape from the well, then at any
time, the units of the polymer with $n<n_i(t)$ would have $R(n,t)=0$, and
the segments from $n_i(t)$ to $n_o(t)$ would be spread across the region of
width $w$. We shall refer to $n_o(t)$ as the peeling point. On multiplying
the equation (\ref{two}) by $\frac{\partial R(n,t)}{\partial n}$ and
integrating from $n_i(t)$ to $n_o(t)$, we get

\begin{equation}
\label{three}\frac m2\left( \frac{\partial R(n,t)}{\partial n}\right)
_{n_o(t)}^2-V_0=\zeta \int_{n_i(t)}^{n_o(t)}\frac{\partial R(n,t)}{\partial t%
}\frac{\partial R(n,t)}{\partial n}dn 
\end{equation}
In arriving at the Eq. (\ref{three}) we have used the fact that $\left( 
\frac{\partial R(n,t)}{\partial n}\right) _{n_i(t)}=0$ and $%
V(R(n_i(t)))=-V_0 $. If the width $w$ is small, then we are justified in
neglecting the right hand side of the Eq. (\ref{three}) and then we find
that $\left( \frac{\partial R(n,t)}{\partial n}\right) _{n_o(t)}^2=\frac{2V_0%
}m$, which on taking the square root may be written as

\begin{equation}
\label{four}\left( \frac{\partial R(n,t)}{\partial n}\right) _{n_o(t)}=\frac{%
F_c}m 
\end{equation}
where $F_c=\sqrt{2mV_0}$. (If we picture the process as in Fig. \ref{fig5},
then we need not worry about the case where $F_c=-\sqrt{2mV_0}$). Now, all
that we have to do is to solve the diffusion equation (\ref{two}), outside
the well region, subject to the above boundary condition. Outside the well,
the equation (\ref{two}) becomes the simpler diffusion equation

\begin{equation}
\label{five}\zeta \frac{\partial R(n,t)}{\partial t}=m\frac{\partial ^2R(n,t)%
}{\partial n^2} 
\end{equation}
which is a linear equation. As the range of the potential well is small in
comparison with the total length of the polymer, we can take

\begin{equation}
\label{six}R(n,t)=0\qquad \text{for\qquad }n<n_o(t) 
\end{equation}

Thus we now have a moving boundary value problem: we have to solve the Eq. (%
\ref{five}) for $R(n,t)$, and determine $n_o(t)$, such that the conditions (%
\ref{one1}), (\ref{four}) and (\ref{six}) are satisfied (see Fig. \ref{fig5}
for a pictorial representation of the process). In thus simplifying the
equation, our assumption of neglecting the right hand side of the Eq. (\ref
{three}) is crucial. Physically, what is being done is that near the point
of breaking away from the potential, the segments are assumed to be at
equilibrium - that means the time scale for peeling the polymer away from
the well is slow in comparison with the time scale for establishment of
local equilibrium near the peeling point.

\begin{figure}
\epsfbox[22 345 569 695]{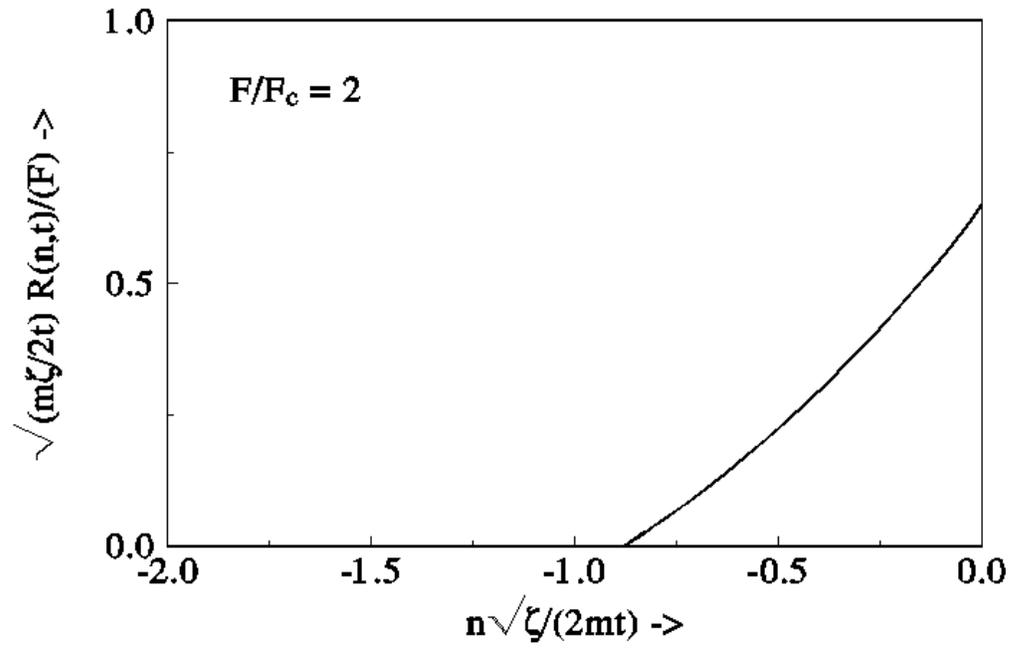}
\caption{Plots of $\sqrt{\frac{m\zeta }{2t}}\frac{R(n,t)}F$ against $n\sqrt{\frac \zeta {2mt}}$.}  \label{fig6}
\end{figure}

Usually, moving boundary value problems are difficult to solve. However, in
this case, it is possible to find similarity solutions to these equations.
For this, we put $R(n,t)=\sqrt{t}\rho (z)$, where $z=n/(\alpha \sqrt{t})$,
with $\alpha =\sqrt{\frac{2m}\zeta }$.

\begin{equation}
\label{seven}\frac{d^2\rho (z)}{dz^2}+z\,\frac{d\rho (z)}{dz}-\rho (z)=0
\end{equation}
Note that $z$ $\epsilon $ $(-\infty ,0)$. The boundary conditions become

\begin{equation}
\label{seven1}\left( \frac{d\rho (z)}{dz}\right) _{z=0}=\frac{\alpha F}m 
\end{equation}

\begin{equation}
\label{seven2}\left( \frac{d\rho (z)}{dz}\right) _{z=zo}=\frac{\alpha F_c}m 
\end{equation}
and

\begin{equation}
\label{seven3}\rho (z_o)=0 
\end{equation}
where $z_o$ is related to $n_o(t)$ by

$$
n_o(t)=z_o\alpha \sqrt{t} 
$$
These equations can be solved to find $\rho (z)$ and $z_o$. The result is:

\begin{equation}
\label{eight}\rho (z)=\frac \alpha {m%
\mathop{\rm Erf}
(\frac{z_0}{\sqrt{2}})}\left[ \left( F_c-F\right) \,\sqrt{\frac 2\pi }\,e^{-%
\frac{z^2}2}+z\,\,\left( \left( F_c-F\right) \,%
\mathop{\rm Erf}
(\frac z{\sqrt{2}})+F%
\mathop{\rm Erf}
(\frac{z_0}{\sqrt{2}})\right) \right] 
\end{equation}
where $z_o(<0)$ is the solution of the equation

\begin{equation}
\label{nine}\,\sqrt{\frac \pi 2}\,%
\mathop{\rm Erf}
(\frac{z_0}{\sqrt{2}})e^{\frac{z_0^2}2}z_0=\frac{\left( F-F_c\right) }{F_c} 
\end{equation}
Note that $z_0$ is a function of $\frac{\left( F-F_c\right) }{F_c}$ only.
Using all the above results, we get:

\begin{equation}
\label{ten}\frac{R(n,t)}F\sqrt{\frac{\zeta m}{2t}}=n\,\sqrt{\frac \zeta {2mt}%
}-\frac{\left( F-F_c\right) \,}{F\,%
\mathop{\rm Erf}
(\frac{z_0}{\sqrt{2}})}\left( \sqrt{\frac 2\pi }e^{-\frac{n^2\,\zeta }{%
4\,m\,t}}+n\,\sqrt{\frac \zeta {2mt}}%
\mathop{\rm Erf}
(\frac 1{\sqrt{2}}n\,\sqrt{\frac \zeta {2mt}})\right) 
\end{equation}
To demonstrate the nature of this solution, we note that the right hand side
is a function of $n\,\sqrt{\frac \zeta {2mt}}$ alone and make a plot of the
left hand side against this variable, in the Fig. \ref{fig6}. The important
conclusion that we can draw from the above is that the point of detachment
of the polymer from the well is given by the equation $n_o(t)=z_0\sqrt{\frac{%
2mt}\zeta }$, where $z_0$ is the solution of the equation (\ref{nine}) and
is $t$ independent. As the left hand side of the equation (\ref{nine}) is
positive, one has a solution for $z_0$ only if $\frac{\left( F-F_c\right) }{%
F_c}>0$ and again this just means the existence of the critical force (a
similar conclusion is arrived at by Bhattacharjee \cite{Bhattacharjee}). His
answer is slightly different, as his analysis takes the noise term too in to
account). The pulling out of the polymer is complete when $n_o(t)=-N$. Hence
the time required to pull out the chain is given by 
$$
t_{po}=\left( \frac N{z_0\alpha }\right) ^2 
$$
For forces slightly exceeding the critical force, one can solve the Eq. (\ref
{nine}) and get $z_0=-\left( \frac{\left( F-F_c\right) }{F_c}\sqrt{\frac 2\pi
}\right) ^{1/2}$. Hence we conclude: $t_{po}\sim N^2$. Further, for small $%
\left( F-F_c\right) $,

\begin{equation}
\label{eleven}t_{po}\sim \left( F-F_c\right) ^{-1}. 
\end{equation}

\section{Conclusions}

\label{conclusions} We now summarize our results and point out the drawbacks
of our analysis. Using the Rouse model, we find that for a polymer trapped
in a potential well, subject to a force at one of its ends, there exists a
critical force $F_c$ and below this critical force, the pulling out is an
activated process, with an activation energy proportional to the number of
segments in the chain. So the pulling out would not happen. For forces
higher than $F_c$, there is no barrier and the process would occur
spontaneously. The dynamics of the pulling out is described by a non-linear
Rouse equation, which is rather difficult to solve. We find an approximate
solution by (a) confining ourselves to the average behavior of the positions
of the segments and (b) introducing a local equilibrium assumption, which
enables us to convert the non-linear equation to a moving boundary value
problem. This moving boundary value problem has a similarity solution and
this leads to the conclusion that the time required to pull the chain out
scales like $N^2$.

Obviously, there is scope for improvement of our analysis. One would like to
perform a more detailed analysis, including the effects of noise. Also, the
effects of hydrodynamic interactions and the excluded volume interactions
have to be included. Inclusion of these may be done by modifying the
equation \ref{one} to

$$
\zeta \frac{\partial R(n,t)}{\partial t}=m\int dn_1h(n-n_1)\frac{\partial
^2R(n_1,t)}{\partial n_1^2}-V^{\prime }(R(n,t))+f(n,t) 
$$
(see the equation (4.50) of Ref. \cite{DE}). In general $h(n)\sim
n^{-\upsilon }$ with $v<1$ (different values can be used for $\nu $ to
account for Zimm dynamics or excluded volume interactions - see equations
(4.49) and (4.69) of Ref. \cite{DE}). With the boundary conditions same as
in our equations (\ref{one2}), (\ref{four}) and (\ref{six}), then one finds
that $t_{po}\sim $ $N^{1+\upsilon }$.

Our results above, however, are only the first few steps in the analysis of
this very interesting problem. The most important next step, obviously, is
to go beyond our mean field description by including the fluctuations. It
seems likely that fluctuations can modify the long time behavior
qualitatively.

\section{Acknowledgments}

I thank Prof. S.M. Bhattacharjee for his comments, Prof. P.L. Sachdev for a
discussion on nonlinear PDE and Ms. Bidisa Das for help with the figures. 

\newpage
\newpage
\newpage
\vspace{6in}
\newpage
\newpage
\newpage
\vspace{6in}

\end{document}